\documentclass[12pt]{article}

\usepackage{graphicx}
\begin{document}

\newcommand{\siml}{\stackrel{<}{\sim}}
\newcommand{\simg}{\stackrel{>}{\sim}}
\newcommand{\lleq}{\stackrel{<}{=}}

\baselineskip=1.333\baselineskip


%
\begin{center}
{\large\bf
Stochastic bifurcation
in FitzHugh-Nagumo ensembles
subjected to additive and/or multiplicative noises
} 
\end{center}

\begin{center}
Hideo Hasegawa
\footnote{E-mail address:  hideohasegawa@goo.jp}
\end{center}

\begin{center}
{\it Department of Physics, Tokyo Gakugei University  \\
Koganei, Tokyo 184-8501, Japan}
\end{center}
\begin{center}
({\today})
\end{center}
\thispagestyle{myheadings}

\begin{abstract}
We have studied the dynamical properties
of finite $N$-unit
FitzHugh-Nagumo (FN) ensembles
subjected to additive and/or multiplicative noises, 
reformulating the augmented moment method (AMM) 
with the Fokker-Planck equation (FPE) method
[H. Hasegawa, J. Phys. Soc. Jpn. {\bf 75}, 033001 (2006)].
In the AMM, original $2N$-dimensional stochastic equations
are transformed to eight-dimensional deterministic ones, and
the dynamics is described in terms of
averages and fluctuations of local and global variables.
The stochastic bifurcation is discussed 
by a linear stability analysis of
the {\it deterministic} AMM equations.
The bifurcation transition diagram
of multiplicative noise is rather different from that
of additive noise: the former has the wider
oscillating region than the latter.
The synchronization in globally coupled FN ensembles
is also investigated.
Results of the AMM are in good agreement
with those of direct simulations (DSs).

\end{abstract}

\noindent
\vspace{0.5cm}

\noindent
\vspace{0.5cm}


%
\noindent
\vspace{0.5cm}

{\rm Telephone: +81-42-329-7482}

{\rm Fax: +81-42-329-7491}

{\rm E-mail: hideohasegawa@goo.jp}

\newpage
\section{INTRODUCTION}

The FitzHugh-Nagumo (FN) model \cite{Fitz61,Nagumo62}
has been widely adopted as a simple model for a wide class
of subjects not only for neural networks
but also for reaction-diffusion chemical systems.
Many studies have been made for the FN model with
single elements \cite{Schurrer91}-\cite{Acebron04}
and globally-coupled ensembles \cite{Acebron04}-\cite{Zaks05}. 
The FN model is usually solved by direct simulation (DS)
or the Fokker-Planck equation (FPE) method.
For $N$-unit FN model, DS requires the computational time
which grows as $N^2$ with increasing $N$.
The FPE method leads to $(2N+1)$-dimensional
partial equations to be solved with appropriate
boundary conditions.
A useful semi-analytical method for
stochastic equations has been proposed,
taking account of the first and second moments of 
variables \cite{Rod96}.
Recently we have proposed 
in \cite{Hasegawa03a,Hasegawa04}  
the augmented moment method (AMM)
based on a macroscopic point of view.
In the AMM, we describe the properties of the stochastic
ensembles in terms of a fairly small number
of variables, paying our attention to their global 
behavior.
For the $N$-unit stochastic systems, each of which is 
described by $K$ variables,
$KN$-dimensional stochastic equations are transformed
to $N_{eq}$-dimensional {\it deterministic} equations in the AMM
where $N_{eq}=K(K+2)$ independent of $N$.
This figure is, for example, $N_{eq}=3$ for the Langevin model ($K=1$)
and $N_{eq}=8$ for the FN model ($K=2$).
The AMM has been successfully
applied to a study on the dynamics of coupled stochastic systems 
described by the Langevin, FN and
Hodgkin-Huxley models subjected to additive noises
with global, local or small-world couplings
(with and without transmission delays)
\cite{Hasegawa03a,Hasegawa04,Hasegawa}. 

In recent years, the noise-induced phenomena such as
stochastic resonance, noised-induced ordered state
and noised-induced bifurcation
have been extensively studied.
Interesting phenomena caused by additive and multiplicative 
noises have been intensively investigated
(for a recent review, see Refs. \cite{Lindner04,Munoz04}:
related references therein). 
It has been realized that
the properties of multiplicative noises
are different from those of additive noises
in some respects as follows.
(1) Multiplicative noise induces the transition,
creating an ordered state, while additive noise
is against the ordering \cite{Broeck94}-\cite{Munoz05}.
(2) Although the probability distribution in Langevin systems
subjected to additive Gaussian noise follows the Gaussian,
multiplicative Gaussian noise 
generally yields non-Gaussian distribution
\cite{Sakaguchi01}-\cite{Hasegawa05b}.
(3) The scaling relation of the effective 
strength: $\beta(N)=\beta(1)/\sqrt{N}$ 
valid for additive noise 
is not applicable for multiplicative noise:
$\alpha(N) \neq \alpha(1)/\sqrt{N}$, where $\alpha(N)$ and $\beta(N)$
denote effective strengths of multiplicative
and additive noises, respectively, of $N$-unit systems
\cite{Hasegawa06}.

In order to show the above item (3), the present author has
adopted the AMM for the Langevin model
in a recent paper \cite{Hasegawa06}.
The AMM was originally
developed by expanding variables
around their mean values in stochastic equations
to obtain the second-order moments both for
local and global variables
\cite{Hasegawa03a}.  
To extend the applicability of 
the AMM to stochastic systems
including multiplicative noises,
we have reformulated it for the Langevin model 
with the use of the FPE
\cite{Hasegawa06,Hasegawa06b}.
It has been pointed out \cite{Hasegawa06}
that a naive approximation of the scaling relation 
for multiplicative noise: $\alpha(N)=\alpha(1)/\sqrt{N}$, 
as adopted in \cite{Munoz05},
leads to the result
which is in disagreement with
that of DSs.

It is doubly difficult to study analytically 
the {\it dynamical} properties of 
stochastic systems with {\it finite} populations.
Most of analytical theories having been proposed so far 
are limited to infinite systems. Usually we solve the FPE 
for $N=\infty$ ensembles by using
the mean-field and diffusion approximations
to get the stationary probability distribution.
For a study of dynamics, we have to obtain the 
instantaneous probability distribution from
the partial differential equations (DEs) within the FPE,
which is difficult even for $N=\infty$.
Recently, the time-dependent probability distribution
is treated with a series expansion 
of the Hermite polynomials, with which
dynamics of $N=\infty$ stochastic systems
is expressed by 
the time-dependent expansion coefficients
\cite{Acebron04}.
In the AMM, equations of motions for $N_{eq}$ moments 
which have clear physical meanings
may describe the dynamics of stochastic systems
with finite $N$.

In this paper, we will study effects of additive and/or
multiplicative noises on 
the dynamical properties of the FN model. 
Although effects of additive noise on the FN model 
have been extensively investigated 
\cite{Schurrer91}-\cite{Zaks05}, 
there have been no such studies on the
effect of multiplicative noise, 
as far as the author is concerned.
We are interested in the stochastic bifurcation, which
is one of interesting phenomena induced by noise
(Refs. \cite{Nama90,Schenk96}, related references therein).
The theory on stochastic bifurcation is still in its infancy.
Indeed, there is no stringent definition of the stochastic
bifurcation.
At the moment, two kinds of definitions have been proposed:
(i) one is based on a sudden change in the stationary
probability distribution, and (ii) the other is based on
a sudden change in the sign of the largest Lyapunov index.
Unfortunately these two definitions do
not necessarily yield the same result.
The bifurcation of the single \cite{Tanabe01}
and ensemble FN model \cite{Acebron04,Zaks05}
subjected to additive noise has been recently discussed.
Based on the second-order moment
method, the bifurcation analysis has been made 
for globally-coupled FN model in Ref. \cite{Zaks05}, 
where dynamics of fast variables
is separated from and projected to that of slow variables
subjected to additive noise.
We will discuss the bifurcation in the FN ensembles
subjected to additive and/or multiplicative noises,
making a linear stability analysis 
to our {\it deterministic} AMM equations.
It is much easier to study the deterministic DEs
than stochastic DEs.

The purpose of the present paper is two folds:
(i) to reformulate AMM for the FN model
subjected to additive and multiplicative noises
with the use of FPE \cite{Hasegawa06,Hasegawa06b}, 
and (ii) to discuss the respective roles of 
additive and multiplicative noises 
on the stochastic bifurcation and synchronization. 
The paper is organized as follows.
In Sec. II, we will apply the AMM
to finite $N$-unit FN ensembles subjected to
additive and multiplicative noises.
With the use of the AMM equations, 
the stochastic bifurcation is discussed in Sec. III. 
Some discussions on
the synchronization are presented in Sec. IV. 
The final Sec. V is devoted to conclusion.


\section{FN neuron ensembles}
\subsection{Description of our model}

We have adopted $N$-unit stochastic systems
described by the FN model
subjected to additive and multiplicative noises.
Dynamics of the coupled ensemble 
is expressed by nonlinear DEs given by 
\begin{eqnarray}
\frac{dx_{i}}{dt} &=& F(x_{i})
- c \:y_{i}
+\alpha\: G(x_i) \eta_i(t)+ \beta \:\xi_i(t)
+I_i^{(c)}(t)+I^{(e)}(t), \\
\frac{dy_{i}}{dt} &=& b \:x_{i} - d \:y_{i}+e,
\hspace{2cm}\mbox{($i=1$ to $N$)}
\end{eqnarray}
with
\begin{equation}
I_i^{(c)}(t)= \frac{J}{Z} \sum_{j\neq i} (x_j-x_i).
\end{equation}
In Eq. (1)-(3),
$F(x)=a_3 x^3+a_2 x^2+a_1 x$,
$a_3=-0.5$, $a_2=0.55$, $a_1=-0.05$,
$b=0.015$, $c=1.0$, $d=0.003$ and $e=0$
\cite{Rod96,Hasegawa03a}: 
$x_i$ and $y_i$ denote the fast and slow variables, respectively:
$G(x)$ is an arbitrary function of $x$:
$I^{(e)}(t)$ stands for an external input:
$J$ expresses the strength of diffusive couplings, $Z=N-1$: 
$\alpha$ and $\beta$ denote magnitudes of
multiplicative and additive noises, respectively,
and $\eta_i(t)$ and $\xi_i(t)$ express zero-mean Gaussian white
noises with correlations given by
\begin{eqnarray}
\langle \eta_i(t)\:\eta_j(t') \rangle
&=& \delta_{ij} \delta(t-t'),\\
\langle \xi_i(t)\:\xi_j(t') \rangle 
&=& \delta_{ij} \delta(t-t'),\\
\langle \eta_i(t)\:\xi_j(t') \rangle &=& 0.
\end{eqnarray}
  
The Fokker-Planck equation $p(\{x_i\},\{y_i\},t)$
is expressed 
in the Stratonovich representation by\cite{Hasegawa06}\cite{Haken83}
\begin{eqnarray}
\frac{\partial}{\partial t} p
&=&-\sum_k \frac{\partial}{\partial x_k}\{ [F(x_k)-cy_k+I_k]p \}
- \sum_k \frac{\partial}{\partial y_k}[(bx_k-dy_k+e)p] \nonumber \\
&+&\left(\frac{\alpha^2}{2}\right) \sum_k \frac{\partial}{\partial x_k}
\{ G(x_k)\:\frac{\partial}{\partial x_k}[G(x_k)\:p] \}
+\sum_k \left( \frac{\beta^2}{2} \right)
\frac{\partial^2 }{\partial x_k^2}\:p,
\end{eqnarray}
where $I_k=I_k^{(c)}+I^{(e)}$.

We are interested also in dynamics of
global variables $X(t)$ and $Y(t)$ defined by
\begin{eqnarray}
X(t)=\frac{1}{N} \sum_i x_i(t), \\
Y(t)=\frac{1}{N} \sum_i y_i(t). 
\end{eqnarray}
The probability of $P(X,Y,t)$ is expressed 
in terms of $p(\{x_i\},\{y_i\},t)$ by
\begin{equation}
P(X,Y,t)= \int \int \: \Pi_idx_i dy_i\:  \:p_i(\{x_i\},\{y_i\},t)
\:\delta \left( X-\frac{1}{N} \sum_i x_i \right) 
\:\delta \left( Y-\frac{1}{N} \sum_i y_i \right).
\end{equation}
Moments of local and global variables
are expressed by
\begin{eqnarray}
\langle x_i^k \:y_i^{\ell} \rangle
&=&
= \int \int d x_i d y_y \;p_i(x_i,y_i,t) \:x_i^k y_i^{\ell}, \\
\langle X^k\:Y^{\ell} \rangle
&=& \int \int dX dY P(X,Y,t) \;X^k Y^{\ell}.
\end{eqnarray}

By using Eqs. (1), (2), (7) and (11), 
we get equations of motions for
means, variances and covariances of local variables by
\begin{eqnarray}
\frac{d \langle x_i \rangle}{dt}&=& \langle F(x_i) \rangle 
-c \langle y_i \rangle
+\frac{\alpha^2}{2} \langle G'(x_i)G(x_i) \rangle, \\
\frac{d \langle y_i \rangle }{dt}
&=&b \langle x_i \rangle -d  \langle y_i \rangle +e, \\
\frac{d \langle x_i x_j \rangle}{dt}
&=& \langle x_i F(x_j) \rangle 
+  \langle x_j F(x_i) \rangle 
-c(\langle x_i y_j \rangle + \langle x_j y_i \rangle)
\nonumber \\
&&+\frac{J}{Z}\sum_k (\langle x_ix_k \rangle
+\langle x_jx_k \rangle  - \langle x_i^2\rangle
-\langle x_j^2 \rangle) \nonumber \\
&&+\frac{\alpha^2}{2} [\langle x_iG'(x_j)G(x_j) \rangle 
+\langle x_jG'(x_i)G(x_i) \rangle ] \nonumber \\
&&+[\alpha^2 \langle G(x_i)^2 \rangle +\beta^2] \delta_{ij}, \\
\frac{d \langle y_i y_j \rangle }{dt}
&=&b (\langle x_iy_j \rangle + \langle x_jy_i \rangle)
-2 d \langle y_i y_j \rangle, \\
\frac{d \langle x_i\:y_j \rangle}{dt}
&=& \langle y_j\:F(x_i) \rangle 
-c \langle y_i y_j \rangle +b \langle x_ix_j \rangle
-d \langle x_i\:y_j \rangle  \nonumber \\
&&+\frac{w}{Z}\sum_k ( \langle x_k\:y_j \rangle -\langle x_i\:y_j \rangle) 
+\frac{\alpha^2}{2} \langle y_j\:G'(x_i)G(x_i) \rangle, 
\end{eqnarray}
where $G'(x)=d G(x)/d x$.
Equations (13)-(17) may be obtainable also with the
use of the Furutsu-Novikov theorem 
\cite{Furutsu63,Novikov65}.

From Eqs. (8), (9) and (12), we get
equations of motions
for variances and covariances of global variables:
\begin{eqnarray}
\frac{d \langle V_{\kappa} \rangle}{dt}
&=& \frac{1}{N} \sum_i \langle v_{\kappa i} \rangle , \\
\frac{d \langle V_{\kappa}\:V_{\lambda} \rangle}{dt}
&=& \frac{1}{N^2}\sum_i\sum_j 
\frac{d \langle v_{\kappa i}\:v_{\lambda j} \rangle}{dt},
\hspace{1cm}\mbox{($\kappa, \gamma=1,2$)}
\end{eqnarray}
where we adopt a convention: 
$v_{1i}=x_i$, $v_{2i}=y_i$, $V_1=X$ and $V_2=Y$.
Equations (13) and (14) were used 
for single ($N=1$) and infinite $(N=\infty)$ FN neurons
subjected only to additive noise ($\alpha=0$)
in the mean-field approximation \cite{Acebron04}.
Equations (13)-(17) were employed
in the moment method for a single FN neuron
subjected to additive noises \cite{Rod96}. 
We will show that Eqs. (18) and (19) play important roles
in discussing finite $N$-unit FN ensembles.

\subsection{AMM equations}

In the AMM \cite{Hasegawa03a}, 
we define eight quantities given by
\begin{eqnarray}
\mu_{\kappa} &=& \langle V_{\kappa} \rangle
= \frac{1}{N} \sum_i \langle v_{\kappa i} \rangle, \\
\gamma_{\kappa,\lambda}
&=&\frac{1}{N} \sum_i \langle (v_{\kappa i}-\mu_{\kappa})
\:(v_{\lambda i}-\mu_{\lambda}) \rangle, \\
\rho_{\kappa,\lambda}
&=& \langle (V_{\kappa}-\mu_{\kappa})\:(V_{\lambda}-\mu_{\lambda}) \rangle, 
\hspace{1cm}\mbox{($\kappa, \lambda=1,2$)}
\end{eqnarray}
with $\gamma_{1,2}=\gamma_{2,1}$ and $\rho_{1,2}=\rho_{2,1}$.
It is noted that $\gamma_{\kappa,\lambda}$ 
expresses the averaged fluctuations
of local variables while
$\rho_{\kappa,\lambda}$ denotes those
of global variables.

Expanding Eqs. (13)-(19) around means of 
$\mu_{\kappa}$ as $v_{\kappa i}=\mu_{\kappa}+\delta v_{\kappa i}$
and retaining the terms of 
$O(\langle \delta v_{\kappa i} \delta v_{\lambda j}\rangle)$,
we get equations of motions for the eight quantities given by
\begin{eqnarray}
\frac{d \mu_1}{dt} &=& f_o+f_2\gamma_{1,1}-c \mu_2
+\frac{\alpha^2}{2}[g_0g_1+3(g_1g_2+g_0g_2)\gamma_{1,1}]
+I^{(e)}, \\
\frac{d \mu_2}{dt}&=& b \mu_1-d \mu_2+e, \\
\frac{d \gamma_{1,1}}{dt} &=& 2(a \gamma_{1,1}-c \gamma_{1,2})
+\frac{2 J N}{Z} (\rho_{1,1}-\gamma_{1,1})
+2\alpha^2 (g_1^2+2g_0g_3) \gamma_{1,1} \nonumber \\
&+& \alpha^2 g_0^2
+\beta^2, \\
\frac{d \gamma_{2,2}}{dt}&=& 2(b\gamma_{1,2}-d\gamma_{2,2}), \\
\frac{d \gamma_{1,2}}{dt} &=& b\gamma_{1,1}
+(a-d)\gamma_{1,2}-c\gamma_{2,2}
+\frac{J N}{Z}(\rho_{1,2}-\gamma_{1,2})
+\frac{\alpha^2}{2}(g_1^2+2g_0g_2)\gamma_{1,2}, \\
\frac{d \rho_{1,1}}{dt} &=& 2(a \rho_{1,1}-c \rho_{1,2})
+2\alpha^2 (g_1^2+2g_0g_2)  \rho_{1,1} 
+ \frac{\alpha^2 g_0^2}{N} 
+\frac{\beta^2}{N}, \\
\frac{d \rho_{2,2}}{dt}&=& 2(b\rho_{1,2}-d\rho_{2,2}), \\
\frac{d \rho_{1,2}}{dt} &=& b\rho_{1,1}
+(a-d)\rho_{1,2}-c\rho_{2,2}
+\frac{\alpha^2}{2}(g_1^2+2g_0g_2)\rho_{1,2}, 
\end{eqnarray}
with
\begin{eqnarray}
a &=& f_1+3 f_3 \gamma_{1,1}, \\
f_{\ell}&=&(1/\ell \:!)F^{(\ell)}(\mu_1), \\
g_{\ell}&=&(1/\ell \:!)G^{(\ell)}(\mu_1).
\end{eqnarray}
%
The original $2N$-dimensional stochastic equations
given by Eqs. (1)-(3)
are transformed to eight-dimensional deterministic
equations. Equations (23)-(30)
with additive noises only ($\alpha=0$)
reduce to those obtained previously 
\cite{Hasegawa03a}.
We note that in the limit of $J=0$,
AMM equations lead to
\begin{eqnarray}
\rho_{\kappa,\lambda} &=& \frac{\gamma_{\kappa,\lambda}}{N},
\hspace{1cm}\mbox{($\kappa, \lambda=1,2)$}
\end{eqnarray}
which is nothing but the central-limit theorem
describing the relation between fluctuations
in local and average variables.
In the limit of $N=1$, we get
$\rho_{\kappa,\lambda} = \gamma_{\kappa,\lambda}$,
by which the AMM equations reduces to the five-dimensional DEs 
for $\mu_1$, $\mu_2$, $\gamma_{1,1}$, $\gamma_{2,2}$
and $\gamma_{1,2}$.

Equations (1)-(3) for $I^{(e)}=\alpha=\beta=0$
have the stationary solution of
$x_i=y_i=0$.
When we assume $G(x)$ for the multiplicative noise given by
\begin{equation}
G(x)=x, 
\end{equation}
the AMM equations are expressed by
\begin{eqnarray}
\frac{d \mu_1}{dt} &=& f_o+f_2\gamma_{1,1}-c \mu_2
+\frac{\alpha^2 \mu_1}{2}+I^{(e)}, \\
\frac{d \mu_2}{dt}&=& b \mu_1-d \mu_2+e, \\
\frac{d \gamma_{1,1}}{dt} &=& 2(a \gamma_{1,1}-c \gamma_{1,2})
+\frac{2J N}{Z} (\rho_{1,1}-\gamma_{1,1})
+2 \alpha^2 \gamma_{1,1} + \alpha^2 \mu_1^2
+\beta^2, \\
\frac{d \gamma_{2,2}}{dt}&=& 2(b\gamma_{1,2}-d\gamma_{2,2}), \\
\frac{d \gamma_{1,2}}{dt} &=& b\gamma_{1,1}
+(a-d)\gamma_{1,2}-c\gamma_{2,2}
+\frac{J N}{Z}(\rho_{1,2}-\gamma_{1,2})
+\frac{\alpha^2\gamma_{1,2}}{2}, \\
\frac{d \rho_{1,1}}{dt} &=& 2(a \rho_{1,1}-c \rho_{1,2})
+2 \alpha^2 \rho_{1,1} + \frac{\alpha^2 \mu_1^2}{N} 
+\frac{\beta^2}{N}, \\
\frac{d \rho_{2,2}}{dt}&=& 2(b\rho_{1,2}-d\rho_{2,2}), \\
\frac{d \rho_{1,2}}{dt} &=& b\rho_{1,1}
+(a-d)\rho_{1,2}-c\rho_{2,2}
+\frac{\alpha^2\rho_{1,2}}{2}.
\end{eqnarray}
The stochastic bifurcation of the AMM equations given by
Eqs. (36)-(43) will be investigated in Sec. 3.
AMM equations for a more general form of
$G(x)=x \mid x \mid^{s-1}$ ($s \geq 0$)
are presented in Appendix A.
Some numerical results for various $s$ values 
will be discussed in Sec. 4 (Fig. 13).

\subsection{Properties of  the AMM}

Contributions from multiplicative noise have
the more complicated $N$ dependence than those 
from additive noise.
Comparing the $\beta^2$ term in $d \gamma_{1,1}/dt$  
of Eq. (25) to that in $d \rho_{1,1}/dt$ of Eq. (28),
we note that the effective strength of additive noise 
of the $N$-unit system, $\beta(N)$,
is scaled by 
\begin{equation}
\beta(N) = \frac{\beta(1)}{\sqrt{N}}.
\end{equation}
In contrast, a comparison between the $\alpha^2$ terms 
in Eq. (25) and (28) yield the two kinds of scalings:
\begin{eqnarray}
\alpha(N) &=& \frac{\alpha(1)}{\sqrt{N}},
\hspace{1cm}\mbox{for $\mu_1$ term}, \\
\alpha(N) &=& \alpha(1),
\hspace{1cm}\mbox{for $\gamma_{1,1}$ and $\rho_{1,1}$ terms}, 
\end{eqnarray}
The relations given by Eqs. (45) and (46) hold
also for $d \gamma_{1,2}/dt$ and $d \rho_{1,2}/dt$
given by Eqs. (27) and (30).
Thus the scaling behavior of the effective strength of
multiplicative noise is quite different
from that of additive noise, as previously pointed out
for Langevin model \cite{Hasegawa06}.
If the relations: $\alpha(N)=\alpha(1)/\sqrt{N}$
and $\beta(N)=\beta(1)/\sqrt{N}$ hold,
the FPE for $P(X,Y,t)$
of the global variables of $X$ and $Y$ in $N$-unit
systems may be expressed by
\begin{eqnarray}
\frac{\partial}{\partial t} P(X,Y,t)
&=&- \frac{\partial}{\partial X}\{ [F(X)-cY+I]P(X,Y,t) \}
-  \frac{\partial}{\partial Y}[(b X-d Y+e)P(X,Y,t)] \nonumber \\
&+& \frac{\alpha^2}{2N} \frac{\partial}{\partial X}
\{ G(x)\:\frac{\partial}{\partial X}[G(X)\:P(X,Y,t)] \}
+ \frac{\beta^2}{2N}\frac{\partial^2 }{\partial X^2}\:P(X,Y,t).
\end{eqnarray}
Unfortunately it is not the case as shown in Eq. (46),
although Eq. (47) may be valid 
in the case of additive noise only
($\alpha=0$) \cite{Acebron04}.

\section{STOCHASTIC BIFURCATION}

\subsection{Additive versus multiplicative noises}

\subsubsection{The case of $N=1$}

In order to get an insight to the AMM,
we first examine the case of single element ($N=1$), for which
AMM equations 
for $\mu_1$, $\mu_2$, $\gamma_{1,1}$, $\gamma_{2,2}$
and $\gamma_{1,2}$ are given by
\begin{eqnarray}
\frac{d \mu_1}{dt} &=& f_o+f_2\gamma_{1,1}-c \mu_2
+\frac{\alpha^2 \mu_1}{2}+I^{(e)}, \\
\frac{d \mu_2}{dt}&=& b \mu_1-d \mu_2+e, \\
\frac{d \gamma_{1,1}}{dt} &=& 2(a \gamma_{1,1}-c \gamma_{1,2})
+2 \alpha^2 \gamma_{1,1}
+ \alpha^2 \mu_1^2
+\beta^2, \\
\frac{d \gamma_{2,2}}{dt}&=& 2(b\gamma_{1,2}-d\gamma_{2,2}), \\
\frac{d \gamma_{1,2}}{dt} &=& b\gamma_{1,1}
+(a-d)\gamma_{1,2}-c\gamma_{2,2}
+\frac{\alpha^2\gamma_{1,2}}{2},
\end{eqnarray}
where $a=f_1+3 f_3 \gamma_{1,1}$.
We have applied a step input given by
\begin{eqnarray}
I^{(e)}(t)&=& A \:\Theta(t-t_{in}),
\end{eqnarray}
where $A=0.1$, $t_{in}=50$ and $\Theta(x)$ denotes
the Heaviside function: $\Theta(x)=1$ for $x \geq 0$
and zero otherwise.
Equations (48)-(52) for $N=1$
have been solved by using the fourth-order Runge-Kutta method
with a time step of 0.01 and with 
zero initial data. 
Direct simulations (DSs) for the $N$-unit FN model given by Eqs. (1)-(3)
have been performed by using the Heun method with
a time step of 0.003 and with the initial data of 
$x_i(0)$ and $y_i(0)$ which are
randomly chosen from $[-0.01, 0.01]$.
Results of DS are averaged over 1000 trials
otherwise noticed.
All quantities are dimensionless.

Figure 1 shows time courses of $\mu_1(t)$,
$\mu_2(t)$, $\gamma_{1,1}(t)$, $\gamma_{2,2}(t)$
and $\gamma_{1,2}(t)$
calculated by AMM (solid curves) and DS (dashed curves)
with $\alpha=0.01$, $\beta=0.0$ ($N=1$).
A single FN neuron fires
when the external input $I^{(e)}(t)$ is applied for 
$t \geq 50$.
By an applied input, $\mu_1$, $\mu_2$,
$\gamma_{1,1}$, $\gamma_{2,2}$ and $\gamma_{1,2}$
show the time-dependent behavior, and they
approach the stationary values at $t > 300$.
Results calculated by AMM are in good agreement with those
of DS.

Time courses of $\mu_1(t)$ [$\gamma_{1,1}(t)$] 
with $\alpha=0.05$, $0.1$,
$0.2$ and $0.5$ are plotted in Fig. 2(a)-(d)
[Fig. 2(e)-2(h)] when an applied input
given by Eq. (53) is applied. 
With increasing $\alpha$, the magnitude of $\gamma_{1,1}(t)$
is much increased, although
the profile of $\mu_1(t)$ is almost the same except for
$\alpha=0.5$.
The time course of $\mu_1(t)$ for $\alpha=0.5$ in the AMM
shows an oscillation which is expected to be due to
the stochastic bifurcation.
Although the result of DS averaged over 1000 trials
shows no oscillation, that of a single trial
clearly shows the oscillation.

As was shown in Fig. 2(d) and 2(h), 
the bifurcation may be induced by strong noise.
In order to discuss the bifurcation,
we have applied a constant input given by
\begin{eqnarray}
I^{(e)}(t)&=& I.
\end{eqnarray}
The stationary equations given by Eqs. (48)-(52) 
with $d \mu_1/dt =0$ {\it et al.}
are solved by the Newton-Raphson method.
Then dynamics given by Eqs. (48)-(52) is solved 
with initial values of the stationary solutions.
Figures 3(a) and 3(b) show the time courses of $\mu_1(t)$
for inputs of $I=0.1$ and $I=0.5$, respectively, with
multiplicative noise ($\alpha=0.1$ and $\beta=0.0$):
solid and dashed curve express the results of the AMM
and DS with a single trail, respectively.
We note that for $I=0.5$,
$\mu_1$ begins to oscillate and its magnitude
is gradually increased while for $I=0.1$,
$\mu_1$ shows no time development. 
This is more clearly seen in the $\mu_1$-$\mu_2$ plots shown
in Fig. 3(b) and 3(d). 
The oscillation for $I=0.5$ is due to the stochastic bifurcation.
Figures 3(e) and 3(f) will be explained later in Sec. 3.1.2.

We have calculated the bifurcation
transition diagrams by making a linear stability analysis 
to the {\it deterministic} AMM equations.
The $5 \times 5$ Jacobian matrix $C$ of
the AMM equations (39)-(43) is expressed 
with a basis of ($\mu_1$, $\mu_2$,
$\gamma_{1,1}$, $\gamma_{2,2}$, $\gamma_{1,2}$) by
\begin{eqnarray}
C=\left[
\begin{array}{ccccc}
f_0'+ f_2' \gamma_{1,1}+\alpha^2/2 
& -c & f_2 & 0 & 0 \\
b & -d & 0  & 0 & 0 \\
2 [(f_1'+3f_3'\gamma_{1,1})\gamma_{1,1}+\alpha^2 \mu_1] 
& 0 & 2(f_1+6 f_3 \gamma_{1,1} +\alpha^2)
& 0 & - 2 c \\
0 & 0 & 0  & -2d & 2 b \\
(f_1'+3f_3' \gamma_{1,1})\gamma_{1,2} & 0 & b + 3 f_3 \gamma_{1,2}  
& -c 
& a-d+\alpha^2/2
\end{array}
\right],
\end{eqnarray}
where $a=f_1+3 f_3 \gamma_{1,1}$ and $f_k'=d f_k/d \mu_1$.
The instability is realized
when any of real parts of five eigenvalues 
in the Jacobian matrix $C$ is positive.
For deterministic FN neuron without noises $(\alpha=\beta=0.0)$, 
the critical condition is given by 
$f_0'-d = 3 \:a_3 \mu_1^2+2 \:a_2 \mu_1 + a_1 - d=0$
for the stationary $\mu_1$, from which
the oscillating state is realized for $0.26 < I < 3.34$.

\vspace{0.5cm}
\noindent
{\bf Multiplicative noise}

In order to obtain the transition diagram, we have performed
calculations of the stationary state and eigenvalues
of its Jacobian matrix $C$,
by sequentially changing a model parameter 
such as $\alpha$, $\beta$, $J$ and $I$.
A calculation of the stationary state
for a given $\alpha$ value, for example, has been 
made by the Newton-Raphon method with initial values which
are given from a calculation for the preceding $\alpha$ value.
Figure 4(a) shows
the $I$-$\alpha$ transition diagram obtained
for multiplicative noise ($\beta=0.0$, $N=1$). 
When changing a parameter as mentioned above,
we have the continuous (second-order) and
discontinuous (first-order) transitions.
Solid curves denote the boundaries of 
the second-order transition between
the oscillating (OSC) and non-oscillating (NONOSC).
As for the first-order transition,
at $I=2.0$, for example, the OSC$\rightarrow$NONOSC transition
takes place at $\alpha=0.11$ when $\alpha$ is increased
from below, while the NONOSC$\rightarrow$OSC transition occurs at
$\alpha=0.04$ when $\alpha$ is decreased from above.
The state for $0.04 < \alpha < 0.11$ with hysteresis
is referred to as the ${\rm OSC}^{'}$ state hereafter.
In the ${\rm OSC}^{'}$ state, we have two stationary solutions,
as will be discussed below.

Dashed curves in Fig. 5(a) show the $I$ dependence of
the maximum real part among five eigenvalues,
$\lambda_m$ (referred to as the {\it maximum index}),
for multiplicative noise ($\alpha=0.1$, $\beta=0.0$).
Two dashed curves for $0.19 < I < 2.29$ express
the results of the two stationary solutions in
the ${\rm OSC}^{'}$ state shown in Fig. 4(a).
The lower, dashed curve crosses the zero line
at four points at $I=0.29$, 1.41, 2.39 and 3.41.
For a comparison, the result for the deterministic
model ($\alpha=\beta=0.0$) is plotted by the chain curve.
Although  the OSC state disappears for
fairly strong multiplicative noises 
of $\alpha > 0.16$ at $2 \siml I \siml 3.3$, 
it persists to strong noises at $0 \siml I \siml 1.3$.

Solid and dashed curves in Fig. 6(a) show $\mu_1$ and $\gamma_{1,1}$,
respectively,
of the stationary state as a function of $I$.
Two dashed curves for $0.19 < I < 2.29$ express
$\gamma_{1,1}$ of the two stationary solutions in the ${\rm OSC}^{'}$ state.
In contrast, there is little difference 
in $\mu_1$ for the two stationary solutions.
We note that $\mu_1$ is nearly proportional to $I$
and that the upper curve of $\gamma_{1,1}$ has a broad peak
centered at $I \sim 2$.
Open and filled circles in Fig. 4(a) denote
two sets of parameters of
$(I, \alpha)=(0.1, 0.1)$ and $(0.5, 0.1)$
adopted for the calculations shown in Figs. 3(a) and (b).

\vspace{0.5cm}
\noindent
{\bf Additive noise}

Figure 4(b) shows the $I$-$\beta$ transition diagram
for additive noise ($\alpha=0.0$, $N=1$). 
Solid curve expresses the boundary 
between the OSC and NONOSC states, and
dashed curve denotes the boundary of the first-order transition
relevant to the ${\rm OSC}^{'}$ state with hysteresis.
When an additive noise is included, the boundary
for the OSC-NONOSC states is much modified.
This is explained for the case of $\beta=0.1$ in Fig. 5(b), 
where the chain curve denotes the maximum index $\lambda_m$
for the case of $\alpha=\beta=0.0$ while
the dashed curve shows
the result for additive noise of $\beta=0.1$ ($\alpha=0.0$).
The chain curve crosses the zero line at $I=0.26$ and 3.34,
while the dashed curve has the zeros at four points 
at $I=0.12$, 0.86, 2.75 and 3.48.
The transition diagram has a strange shape, which is symmetric
with respect to the axis of $I=1.80$. 
The $I$ dependences of $\mu_1$ and $\gamma_{1,1}$ are plotted by
solid and dashed curves, respectively, in Fig. 6(b).
$\gamma_{1,1}$ for additive noise has a broad peak similar
to that for multiplicative noise shown in Fig. 6(a).
It is noted that $\sqrt{\gamma_{1,1}}$ 
expresses the effective width of the probability distribution
of $p(x)=\int \:p(x,y) \:dy$.
We note that $\gamma_{1,1}$ in Fig. 6(a) 
shows a rapid increase at $I \sim 0.2$
where the OSC-NONOSC transition takes place in Fig. 4(a).
Except this case, however, 
there are no abrupt changes in $\gamma_{1,1}$
at the transition.

\subsubsection{The case of finite $N$}

\noindent
{\bf Multiplicative noise}

Figure 3(e) shows the time courses of $\mu_1(t)$
for an input of $I=0.5$ given by Eq. (54) 
applied to $N=100$ FN ensembles with $J=1.0$
subjected to multiplicative noise ($\alpha=0.1$, $\beta=0.0$):
solid and dashed curves express the results of the AMM
and DS with a single trail, respectively.
It shows that $\mu_1$ begins to oscillate and its magnitude
is rapidly increased. 
This is more clearly seen in the $\mu_1$-$\mu_2$ plots shown
in Fig. 3(f). In contrast,
for smaller $I=0.1$, we have not obtained the oscillating solution,
just as in the $N=1$ case shown in Figs. 3(a) and 3(b).

We have performed a linear analysis
for the case of finite $N$, by using
the $8 \times 8$ Jacobian matrix
with the stationary solutions obtained from Eqs. (36)-(43),
expressions for $8 \times8$ Jacobian matrix elements 
being presented in Appendix B.

The $I$-$\alpha$ transition diagram for multiplicative 
noise with $N=100$ and $J=1.0$ is shown in Fig. 4(c).
The region of the OSC state is a little
decreased for a weak $\alpha$ but it is widen for
stronger $\alpha$.
The solid curve in Fig. 5(a) expresses
the $I$ dependence of the maximum index $\lambda_m$ for 
multiplicative noise with $\alpha=0.1$ and $J=1.0$,
which crosses the zero line at two points at
$I=0.21$ and 3.37.

Figure 7(a) shows the $J$-$\alpha$ 
transition diagram for multiplicative noise ($\beta=0.0$)
for $I=3.0$ with $N=100$.
For $\alpha=0.1$, the OSC state is realized even for $J=0.0$. 
For strong multiplicative noise of $\alpha=0.3$,
the OSC state disappears at $J \leq 0.365$.
For $\alpha=0.2$, the OSC state is realized not only
at $J \geq 0.194$ but also at $0.085 \leq J \leq 0.136$,
as shown in the inset of Fig. 7(a).
This implies the re-entrance to the OSC state
from the NONOSC state
when the coupling is decreased from above.

\vspace{0.5cm}
\noindent
{\bf Additive noise}

Figure 4(d) shows 
the $I$-$\beta$ transition diagram for additive noise for 
$J=1.0$ and $N=100$.
The width of the OSC state is gradually decreased
with increasing $\beta$.
Chain and solid curves in Fig. 5(b) show
the $I$ dependences of $\lambda_m$ for $\beta=0.0$
and $\beta=0.1$, respectively, with $J=1.0$ and $N=100$.
The former has the zeros at $I=0.26$ and 3.34
while the latter at $I=0.29$ and 3.32.

Figure 7(b) shows the $J$-$\beta$ transition diagram
for additive noise for $I=3.0$ with $N=100$.
The critical values of $\beta$ are 0.114,
0.221 and 0.265 for $J=0.0$, 0.5 and 1.0,
respectively.
The region of the OSC state is increased
with increasing $J$.

When we compare the bifurcation transition diagrams
for multiplicative noise in Fig. 4(a) and 4(c)
with those for additive noise in Fig. 4(b) and 4(d), 
we note 
that the former is rather different from the latter,
in particular for the $N=1$ case.
When a weak additive noise is added to the $N=1$ model,
the OSC state is slightly increased although for a large noise,
the OSC state disappears.
In contrast, the OSC state persists for multiplicative noise. 
Figures 7(a) and 7(b) show that 
the coupling is beneficial to the OSC state
for both additive and multiplicative noises, as expected.

\subsection{Coexistence of additive and multiplicative noises}

Although we have so far discussed the additive and 
multiplicative noises separately,
we now consider the case where both the noises coexist.
Figure 8(a) shows the $I-\alpha$ transition diagram for
$\beta=0.05$ and $N=1$.
It is similar to the transition diagram shown in Fig. 4(a)
for the case of multiplicative noise only ($\beta=0.0$). 
In Fig. 8(a), the OSC state is completely split 
with a gap even for $\alpha=0.0$.
This is because the OSC state disappears for additive noise
of $\beta=0.05$ even without multiplicative noise
as shown in Fig. 4(b).

In contrast,
Fig. 8(b) shows the $I-\beta$ transition diagram for
$\alpha=0.1$ and $N=1$.
We have the ${\rm OSC}^{'}$ state for
weak $\beta$ at $1.30 < I < 2.38$.
This is related to the fact that the ${\rm OSC}^{'}$ state
exists for $\alpha=0.1$ and $\beta=0.0$ in Fig. 4(a).

Figure 9 shows the $\alpha$-$\beta$ transition diagram 
for the OSC and NONOSC states with $I=1.0$ and $I=3.0$
($N=1$).
In the case of $I=1.0$, the boundary 
between the OSC and NONOSC states
extends to large $\alpha$,
reflecting the behavior shown in Figs. 4(a) and 8(a).
We have tried to fit the calculated boundaries by
a simple expression of $\beta=c \sqrt{1-(\alpha/d)^2}$.
Dashed curves in Fig. 9 express the results
with $c=0.084$ and $d=0.32$ for $I=1.0$
and with $c=0.121$ and $d=0.158$ for $I=3.0$.
An agreement between the solid and dashed curves 
in the case of $I=3.0$
is better than that in the case of $I=1.0$. 


\section{DISCUSSION}


It is interesting to study the synchronization 
in FN ensembles with noises.
In order to quantitatively discuss the
synchronization in the ensemble, 
we first consider the quantity given by
\cite{Hasegawa03a}
\begin{equation}
R(t)=\frac{1}{N^2} \sum_{i j} \langle [x_i(t)-x_j(t)]^2 \rangle
=2 [\gamma_{1,1}(t)-\rho_{1,1}(t)].
\end{equation}
When all neurons are in the completely synchronous state,
we get $x_{i}(t)=X(t)$ for all $i$, and 
then $R(t)=0$ in Eq. (56).
On the contrary, in the asynchronous state, we get 
$R(t)=2(1-1/N)\gamma_{1,1} \equiv R_0(t)$
because $\rho_{1,1}=\gamma_{1,1}/N$ [Eq. (34)].
We modify $R(t)$ such that the synchronization is
scaled between the zero and unity, defining
the synchronization ratio
given by \cite{Hasegawa03a}
\begin{equation}
S(t) =1-\frac{R(t)}{R_0(t)}
= \left[ \frac{N\rho_{1,1}(t)/\gamma_{1,1}(t)-1}{N-1} \right],
\end{equation}
which is 0 and 1 for completely asynchronous ($R=R_0$)  
and synchronous states ($R=0$), respectively.
As will be shown below, $S(t)$ depends not only on model parameters
such as $J$ and $N$ but also the type of noises
($\alpha$ and $\beta$).


The calculations have been performed 
for an external input $I^{(e)}(t)$ given by Eq. (53).
DS calculations have been made 
with 20 trials. 

Figures 10(a)-(d) show time courses of $\mu_1(t)$,
$\gamma_{1,1}(t)$, $\rho_{1,1}(t)$ and $S(t)$
for multiplicative noise
($\alpha=0.01$, $\beta=0.0$) with $J=1.0$ and $N=100$.
In contrast, Figs. 10(e)-(h) express the result
for additive noise ($\alpha=0.01$, $\beta=0.0$).
When the external input
is applied at $t=50$,  
FN neurons fire, and $\gamma_{1,1}(t)$, $\rho_{1,1}(t)$ 
and $S(t)$ develop.
The time dependence of $\mu_1(t)$ for multiplicative noise
in Fig. 10(a) is almost the same as that for additive noises
shown in Fig. 10(e).
From a comparison between Figs. 10(b) and 10(f),
we note, however, that 
$\gamma_{1,1}$ for multiplicative noise
is considerably different from that 
for additive noise.
This is true also for $\rho_{1,1}$ shown in Figs. 10(c)
and 9(g).
The time course of $S(t)$ is calculated with the use of
Eq. (57) with $\gamma_{1,1}$ and $\rho_{1,1}$ shown 
in Figs. 10(b) and 10(c)
[Figs. 10(f) and 10(g)].
Reflecting the differences in $\gamma_{1,1}$
and $\rho_{1,1}$,
the result of $S(t)$ for multiplicative noise 
in Fig. 10(d) is quite different from that for additive noise
in Fig. 10(h),
although at $t > 400$, both $S(t)$
approach the same value of $S=0.24$.

We may apply also spike train and sinusoidal inputs
to the systems, whose results will be discussed 
in the followings.
Figures 11(a) and 11(b) show time courses of
$\mu_1(t)$ and $S(t)$ 
in the case of multiplicative noise
($\alpha=0.01$, $\beta=0.0$) with $J=1.0$ and $N=100$,
when a spike train input 
given by
\begin{eqnarray}
I^{(e)}(t)&=& A \sum_n \Theta(t-t_{n})\Theta(t_n+T_w-t),
\end{eqnarray}
with $A=0.1$, $T_n=50+100 (n-1)$ and $T_w=10$ is applied,
an input being plotted at the bottom of Fig. 11(a).
In order to understand the relation between 
$\mu_1$ and $S$, we depict the $S-(d\mu_1/dt)$ plot
shown by solid curves in Fig. 11(c),
where initial data of $S$ and $d\mu_1/dt $ 
at $t < 350$ are neglected.
We note that $S$ is large for $d \mu_1/dt \sim -0.15$
and $0 < d \mu_1/dt < 0.1$.
In contrast, we show by the dashed curve in Fig. 11(c),
a similar plot in the case of additive noise
($\alpha=0.0$, $\beta=0.01$).
The trend of multiplicative noise is similar to
that of additive noise for $d \mu_1/dt < 0$
but quite different for $d \mu_1/dt > 0$.

Figures 12(a) and 12(b) show time courses of
$\mu_1(t)$ and $S(t)$ 
in the case of multiplicative noise 
($\alpha=0.01$, $\beta=0.0$
with $J=1.0$ and $N=100$,
when a sinusoidal input given by
\begin{eqnarray}
I^{(e)}(t)&=& A \:\Theta(t-t_b)
\left[ 1-\cos\left( \frac{2 \pi (t-t_b)}{T_p}\right) \right],
\end{eqnarray}
with $A=0.1$, $t_b=50$ and $T_p=100$ is applied,
input being plotted at the bottom of Fig. 12(a).
The solid curve of Fig. 12(c) expresses
the $S-(d\mu_1/dt)$ plot, where initial data 
of $S$ and $d \mu_1/d t$ at $t < 350$ are neglected.
In contrast, the dashed curve in Fig. 12(c) shows
the $S-(d\mu_1/dt)$ plot in the case of additive noise
($\alpha=0.0$, $\beta=0.01$).
It is noted that the behavior of
the $S-(d\mu_1/dt)$ plot of multiplicative noise
is similar to that of additive noises 
for $d \mu_1/dt < 0$ but rather different
for $d \mu_1/dt >$.
This is consistent with the result shown in Fig. 11(c)
for spike train.
It is interesting that the synchronization $S$ seems to
correlate with $\mid d \mu_1/dt \mid$:
$S$ becomes larger for a larger $\mid d \mu_1/dt \mid$.

We have discussed the stochastic bifurcation 
and the synchronization
in FN ensembles subjected to additive and/or 
multiplicative noises.
Our calculations have shown that
effects of multiplicative noise are rather different from
those of additive noise.
This may be understood by a simple analysis as follows.
Equation (48) is rewritten as
\begin{eqnarray}
\frac{d \mu_1}{dt} &=& a_3 \mu_1^3 + a_2 \mu_1^2 + a_1 \mu_1
- c\mu_2 +(3 a_3 \mu_1 + a_2) \gamma_{1,1} 
+\frac{\alpha^2}{2} \mu_1 + I.
\end{eqnarray}
For weak noises, we get
\begin{equation}
\gamma_{1,1} \sim D (\alpha^2 \mu_1^2 + \beta^2),
\end{equation}
where $D$ is the coefficient to be determined from Eqs. (48)-(52)
but its explicit form is not necessary for our discussion.
Substituting Eq. (61) to Eq. (60), we get
\begin{eqnarray}
\frac{d \mu_1}{dt} &=& a_3' \mu_1^3 + a_2' \mu_1^2 + a_1' \mu_1
- c\mu_2 + I',
\end{eqnarray}
with
\begin{eqnarray}
a_3' &=& a_3(1+ 3 D \alpha^2), \\
a_2' &=& a_2 (1+ D \alpha^2), \\
a_1' &=& a_1 + \frac{\alpha^2}{2} + 3 D a_3 \beta^2, \\
I' &=& I + D a_2 \beta^2.
\end{eqnarray}
The dynamics of $\mu_1$ and $\mu_2$ is effectively
determined by Eqs. (49) and (62).
Results for deterministic FN model are given by
setting $\alpha=\beta=0.0$ in Eqs. (62)-(66). 
We note in Eqs. (63)-(66) that additive noise modifies  
constant and linear terms, whereas multiplicative noise
changes the linear, quadratic and cubic terms. 
These differences yield the difference in the stochastic
bifurcations for additive and multiplicative noises.

The bifurcation diagram of a single FN model
with additive noise is discussed 
in Refs. \cite{Tuckwell98,Tanabe01,Acebron04}.
It has been shown that the probability distribution
of single FN model obeys Gaussian for weak additive noise
while for strong noise, it shows a deviation from
Gaussian with the bimodal structure \cite{Tanabe01,Zaks03}.
The transition from the Gaussian to non-Gaussian distribution
is considered to show the stochastic bifurcation.
However, a change in the form of probability density
is generally gradual when a parameter of the model is changed.
By employing the second-order moment method, 
Tanabe and Pakdaman \cite{Tanabe01}
have obtained the bifurcation diagram 
showing a critical current $I_c$
of a single FN model with additive noise,
which approaches $I_c$ for the deterministic FN model
with decreasing the strength of additive noise.
Our transition diagram shown in Fig. 4(b) is different from
their transition diagram
(Fig. 4 of Ref. \cite{Tanabe01}) in which
the OSC state exists even for a strong additive noise.
By solving the FPE by an expansion of the Hermite polynomials,
Acebr\'{o}n, Bulsara and Rappel \cite{Acebron04} have shown
that the stochastic bifurcation cannot be obtained
for a single FN model 
against our result showing the bifurcation 
for $N=1$ [Fig. 4(b)].
Recently, stochastic bifurcations 
in globally coupled ($N=\infty$) ensembles
subjected to additive noise have been discussed 
in Refs. \cite{Acebron04,Zaks03,Zaks05}.
It has been shown \cite{Acebron04} 
that contrary to a single element,
the bifurcation occurs in globally coupled ensembles
as the noise strength is increased. 
With the use of the moment method,
Zaks, Sailer, Schimansky-Geier and Neiman  
\cite{Zaks05} have discussed the bifurcation
of coupled ($N= \infty$) FN model
in which slow variables ($y_i$) are subjected to additive noise
whereas in our study, fast variables ($x_i$)
are subjected to additive and/or multiplicative
noises [Eqs. (1) and (2)].
With increasing the noise intensity,
mean field shows a transition from a steady equilibrium to
global oscillations, and then back to another equilibrium 
for sufficiently strong noise \cite{Zaks05}.

In the conventional moment (or cumulant) approach, 
the Gaussian distribution is assumed for calculations of
first and second moments, and the moment method 
is considered to lose its validity
for the non-Gaussian distribution 
\cite{Rod96}-\cite{Tanabe01},\cite{Zaks05}. 
Our reformulation of the AMM with FPE has revealed
that the moment method is free from the Gaussian approximation
and that it is valid also for non-Gaussian distribution.
Indeed, we have shown in Refs. \cite{Hasegawa06,Hasegawa06b}
that the AMM can be well applied to
the Langevin model with multiplicative Gaussian noise
although its probability
distribution generally follows the non-Gaussian
\cite{Sakaguchi01,Anten02}. 
The moment approach is expected
to have the wider applicability
than having been considered so far.

The transition diagrams shown in Figs. 4(a), 4(c) and
4(e) for multiplicative noise have the asymmetric, peculiar
structure, which arises from the assumed form 
of $G(x)=x$ in Eq. (35).
If we alternatively assume $G(x)=1$, 'multiplicative noise'
reduces to additive noise, and its transition diagrams
are given by  Figs. 4(b), 4(d) and 4(f).
Thus the structure of the bifurcation transition diagram
depends on the adopted form of $G(x)$.
Some preliminary calculations have been made 
with the use of a form of
$G(x)=x \mid x \mid^{s-1}$ by changing
the index $s$: the relevant AMM equations are
presented in Appendix A.
%

Figure 13(a) shows time courses of $S(t)$ for various $s$
with $\alpha=0.01$, $\beta=0.0$ and $N=100$.
At a glance, the overall behavior of $S(t)$ seems almost
independent of $s$. 
We note, however, that $\gamma_{1,1}$ and $\rho_{1,1}$, which are
plotted in Figs. 13(b) and 13(c), respectively, much depend
on the $s$ value.
It has been shown that in the Langevin model, the
stationary distribution shows much variety
depending on the index $s$ in $G(x)=x \mid x \mid^{s-1}$ 
\cite{Hasegawa06,Hasegawa06b}.
This is expected to be true also for the distribution $p(x,y)$
in the FN model with multiplicative noises.
It would be interesting to investigate the stochastic
bifurcation by changing the index $s$, whose study 
is under consideration.

\section{CONCLUSION}

We have studied effects of additive and multiplicative noises
in single elements and globally-coupled ensembles
described by the FN model, by employing
AMM reformulated with the use of
FPE \cite{Hasegawa06,Hasegawa03a}.
The stochastic bifurcation has been examined
by a linear analysis of the deterministic AMM equations.
The properties of the multiplicative noise in FN neuron ensembles
is summarized as follows.

\noindent
(a) the effect of multiplicative noise
on the stochastic bifurcation is different from that 
of additive noise, and
 
\noindent
(b) the effect of multiplicative noise on the 
synchronization is more significant than that
of additive noise.

\noindent
The item (a) is consistent with the results obtained for
other nonlinear systems such as Duffing-Van der Pol model 
\cite{Schenk96,He04}. The effect of noise depends
on the type of noises, 
and it is also model dependent \cite{Leung98}.
The item (b) is similar to the item (1) of
an ordered state created by multiplicative noises 
\cite{Broeck94}-\cite{Munoz05} mentioned in the introduction.

A disadvantage of our AMM is that its applicability 
is limited to weak-noise cases. 
In physics, there are many approximate methods 
which are valid in some limits but which provide us with
clear physical picture beyond these limits.
One of such examples is 
the random-phase approximation (RPA)
which has been widely employed in solid-state physics. 
The RPA is valid in the limit of weak interactions.
The RPA is, however, used for larger interactions 
leading to the divergence in the response function,
which signifies the occurrence of excitations such as spin waves.
We expect that the AMM may be such an approximate method,
yielding meaningful qualitative result even for strong noises.

On the contrary, an advantage of the AMM is that we can easily discuss
dynamical properties of the finite $N$-unit stochastic systems.
For $N$-unit FN neuronal ensembles,
the AMM yields the eight-dimensional ordinary DEs, while
DS and FPE require
the $2 N$-dimensional stochastic DEs and
the $(2 N+1)$-dimensional partial DEs, respectively.
Furthermore the calculation of the AMM is much faster than DSs: 
for example, it is about 2000 times faster than DS calculations 
with 100 trials for 100-unit FN ensembles.
We hope that the AMM may be applied to a wide class of 
coupled stochastic systems
subjected to additive and/or multiplicative noises.

\section*{Acknowledgements}
This work is partly supported by
a Grant-in-Aid for Scientific Research from the Japanese 
Ministry of Education, Culture, Sports, Science and Technology.  


\vspace{1cm}

\noindent
{\Large\bf Appendix A: 
AMM equations for $G(x)=x \mid x \mid^{s-1}$}

In the case of $G(x)$ given by
\begin{eqnarray}
G(x)&=& x \mid x \mid^{s-1}
\hspace{1cm}\mbox{($s \geq 0$)},
\hspace{4cm} \mbox{(A1)} \nonumber
\end{eqnarray}
AMM equations (51)-(58) become
\begin{eqnarray}
\frac{d \mu_1}{dt} &=& f_o+f_2\gamma_{1,1}-c \mu_2
+\frac{1}{2}\alpha^2 s \mu_1 \mid \mu_1\mid^{2s-2}
[1+(s-1)(2s-1) \mid \mu_1 \mid^{-2}\gamma_{1,1}] \nonumber \\
&+&I^{(e)}, 
\hspace{9,5cm} \mbox{(A2)} \nonumber
\\
\frac{d \mu_2}{dt}&=& b \mu_1-d \mu_2+e, 
\hspace{7.5cm} \mbox{(A3)} \nonumber
\\
\frac{d \gamma_{1,1}}{dt} &=& 2(a \gamma_{1,1}-c \gamma_{1,2})
+\frac{2J N}{Z} (\rho_{1,1}-\gamma_{1,1})
+2\alpha^2 s(2s-1) \mid \mu_1 \mid^{2s-2} \gamma_{1,1} \nonumber \\
&+& \alpha^2 \mid \mu_1 \mid^{2s}
+\beta^2, 
\hspace{7.5cm} \mbox{(A4)} \nonumber
\\
\frac{d \gamma_{2,2}}{dt}&=& 2(b\gamma_{1,2}-d\gamma_{2,2}), 
\hspace{7.5cm} \mbox{(A5)} \nonumber
\\
\frac{d \gamma_{1,2}}{dt} &=& b\gamma_{1,1}
+(a-d)\gamma_{1,2}-c\gamma_{2,2}
+\frac{J N}{Z}(\rho_{1,2}-\gamma_{1,2}) \nonumber \\
&+& \frac{1}{2}\alpha^2 s(2s-1) \mid \mu_1 \mid^{2s-2} \gamma_{1,2}, 
\hspace{5.5cm} \mbox{(A6)} \nonumber
\\
\frac{d \rho_{1,1}}{dt} &=& 2(a \rho_{1,1}-c \rho_{1,2})
+2\alpha^2 s(2s-1) \mu_1^{2s-2} \rho_{1,1} 
+ \frac{\alpha^2 \mid \mu_1 \mid^{2s}}{N} 
+\frac{\beta^2}{N}, 
\hspace{1cm} \mbox{(A7)} \nonumber
\\
\frac{d \rho_{2,2}}{dt}&=& 2(b\rho_{1,2}-d\rho_{2,2}), 
\hspace{8cm} \mbox{(A8)} \nonumber
\\
\frac{d \rho_{1,2}}{dt} &=& b\rho_{1,1}
+(a-d)\rho_{1,2}-c\rho_{2,2}
+\frac{1}{2}\alpha^2 s(2s-1) \mid \mu_1 \mid^{2s-2} \rho_{1,2}.
\hspace{1cm} \mbox{(A9)} \nonumber 
\end{eqnarray}
Some numerical examples of $\gamma_{11}(t)$ {\it et al.}
for various values of the index $s$ 
are shown in Fig. 13.

\vspace{1cm}
\noindent
{\Large\bf Appendix B: Jacobian matrix of AMM equations}

For an analysis of the stochastic bifurcation
of finite $N$-unit FN ensembles, 
we need the $8 \times 8$ Jacobian matrix $C$ 
expressed with a basis of ($\mu_1$, $\mu_2$, 
$\gamma_{1,1}$, $\gamma_{2,2}$, $\gamma_{1,2}$,
$\rho_{1,1}$, $\rho_{2,2}$, $\rho_{1,2}$), by
\begin{eqnarray}
C&=&\left[
\begin{array}{cccc}
f_0'+ f_2' \gamma_{1,1}+\alpha^2/2 & -c & f_2 & 0  \\
b & -d & 0  & 0  \\
2[(f_1'+3 f_3'\gamma_{1,1}) \gamma_{1,1}+ \alpha^2 \mu_1]
& 0 & 2(f_1+6 f_3 \gamma_{1,1} +\alpha^2)- 2JN/Z  & 0  \\
0 & 0 & 0  & -2d \\
(f_1'+3 f_3' \gamma_{1,1} )\gamma_{1,2} & 0 
& b + 3 f_3 \gamma_{1,2}  & -c \\
2 (f_1'+3 f_3' \gamma_{1,1})\rho_{1,1}+2 \alpha^2 \mu_1/N 
& 0 & 6 f_3 \rho_{1,1} & 0 \\
0 & 0 & 0 & 0 \\
(f_1'+3 f_3' \gamma_{1,1} )\rho_{1,2} & 0 
& 3 f_3 \rho_{1,2} & 0 
\end{array}
\right] \nonumber \\
\nonumber \\ \nonumber \\
&&\left[
\begin{array}{cccc}
0 & 0 & 0 & 0  \\
0 & 0 & 0  & 0  \\
- 2 c & 2JN/Z & 0 & 0  \\
2b & 0 & 0  & 0 \\
a -d+\alpha^2/2 -JN/Z & 0 
& 0  & JN/Z \\
0 & 2 (a+ \alpha^2) & 0 & - 2 c \\
0 & 0 &  -2d & 2 b \\
0 & b & -c & a-d+\alpha^2/2
\end{array}
\right],
\hspace{2cm} \mbox{(B1)} \nonumber
\end{eqnarray}
where $a=f_1+3 f_3 \gamma_{1,1}$ and $f_k'=d f_k/d \mu_1$, 
the (1-4)-column and (5-8)-column elements
being separately expressed for a convenience of printing.

\newpage


\end{document}